\documentclass[aps,prb,preprint,superscriptaddress]{revtex4-1}

\usepackage{graphicx}

\graphicspath{{c:/Users/Elizabeth/Documents/Work/2013/CryostatPhaseDiagram_Paper/Figures/Final_Revisions/}}
\usepackage{mathtools}
\usepackage{gensymb,textcomp}
\begin{document}

\title{Substrate-induced microstructure effects on the dynamics of the photo-induced Metal-insulator transition in VO$_2$ thin films}
\author{E.Radue}
\email[Corresponding author: ]{elradue@email.wm.edu}
\affiliation{Department of Physics \ College of William and Mary \ Williamsburg VA, 23188}
\author{L. Wang}
\affiliation{Department of Physics \ College of William and Mary \ Williamsburg VA, 23188}
\author{S.Kittiwatanakul}
\affiliation{Department of Physics \ University of Virginia \ Charlottesville VA, 22904}
\author{J. Lu}
\affiliation{Department of Material Science \ University of Virginia \ Charlottesville VA, 22904}
\author{S.A. Wolf}
\affiliation{Department of Physics \ University of Virginia \ Charlottesville VA, 22904}
\author{E. Rossi}
\affiliation{Department of Physics \ College of William and Mary \ Williamsburg VA, 23188}
\author{R.A. Lukaszew}
\affiliation{Department of Physics \ College of William and Mary \ Williamsburg VA, 23188}
\author{I. Novikova}
\affiliation{Department of Physics \ College of William and Mary \ Williamsburg VA, 23188}
\begin{abstract}
We investigate the differences in the dynamics of the ultrafast photo-induced metal-insulator transition (MIT) of two VO$_2$ thin films deposited on
 different substrates, TiO$_2$ and Al$_2$O$_3$, and in particular the temperature dependence of the threshold laser fluence values required to induce various MIT stages in a wide range of sample temperatures (150 K - 320 K).  We identified that, although the general pattern of MIT evolution was similar for the two samples, there were several differences.  Most notably, the threshold values of laser fluence required to reach the transition to a fully metallic phase in the VO$_2$ film on the TiO$_2$ substrate were nearly constant in the range of temperatures considered, whereas the VO$_2$/Al$_2$O$_3$ sample showed clear temperature dependence.  Our analysis qualitatively connects such behavior to the structural differences in the two VO$_2$ films. 
\end{abstract}
\maketitle

\section{INTRODUCTION}
Many applications can benefit from the ability to controllably change the electrical and/or optical properties of some materials that can undergo a metal-to-insulator phase transition (MIT). Vanadium dioxide VO$_2$ has been a model material for many such studies, as it undergoes a reversible thermally-induced MIT at a convenient transition temperature (T$_c$=340 K for bulk VO$_2$)\cite{Zylbersztejn1975a,Morin1959}. The MIT in VO$_2$ can also be induced at sub-picosecond timescales using ultrafast optical pulses\cite{Cavalleri1999a}, and also by applying sufficiently strong electric fields.\cite{Gopalakrishnan2009a}

\begin{figure}[!htbp]
\centering
\includegraphics[width=\textwidth]{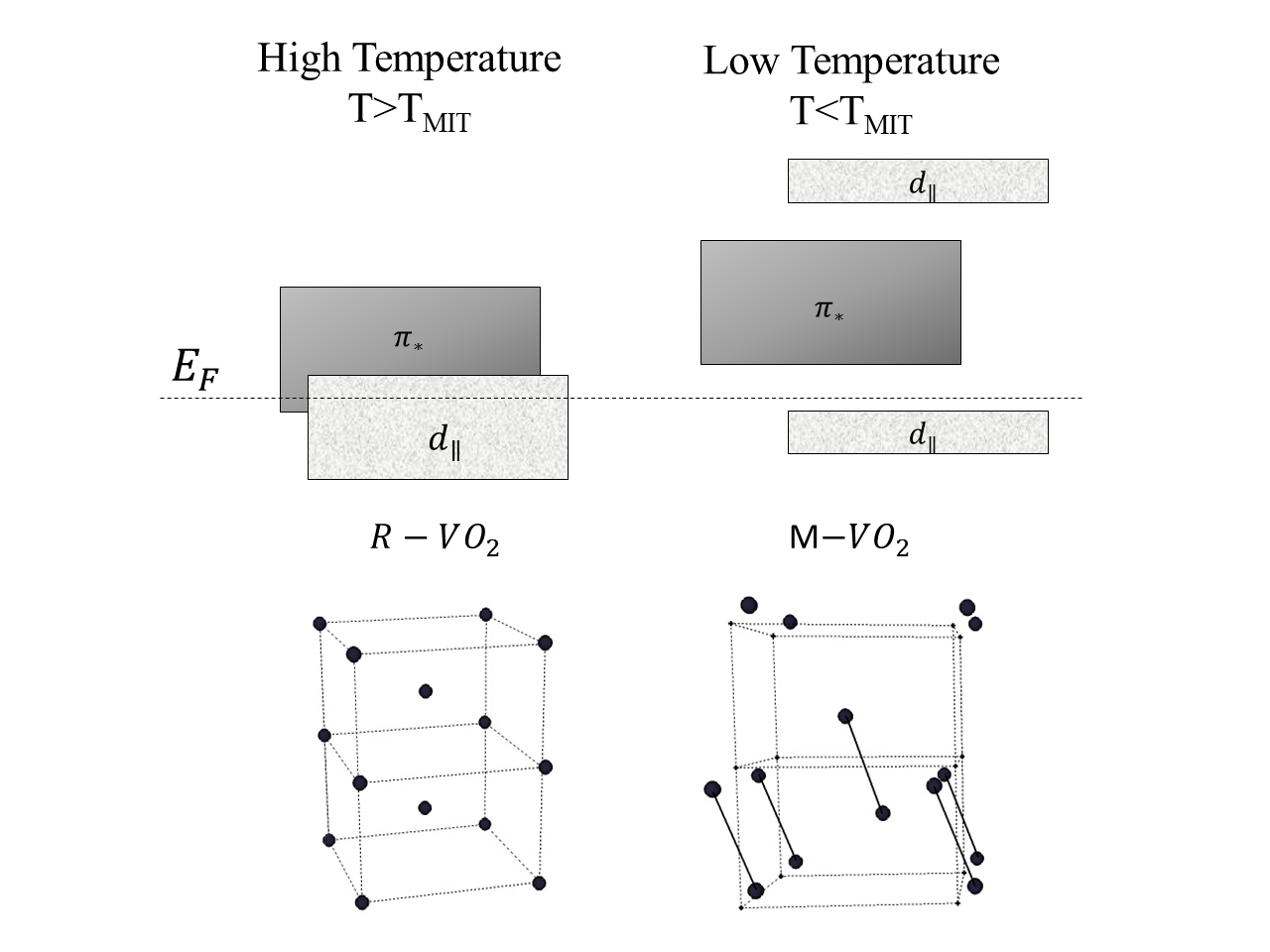}
\caption{Schematics of the VO$_2$ lattice structure and electron energy band structure below and above the critical MIT temperature. High-temperature metallic phase corresponds to the tetragonal (or rutile) lattice structure. During the thermally induced phase transition the V$^{4+}$ ions dimerize along the c-axis of the tetragonal phase, breaking the symmetry and forming monoclinic lattice cells. At the same time the $d$ band splits into two different bands, while the $\pi*$ band shifts above the Fermi level E$_F$, drastically decreasing electron conductivity.\cite{Goodenough1971}}
\label{bandandlattice}
\end{figure}

VO$_2$ belongs to a class of highly-correlated materials, in which the electrical properties depend on strong electron-electron interactions. The MIT is currently understood to be due to the interplay of strong electron-electron interactions and a change of the VO$_2$ crystal structure.\cite{Aetukuri2013,Imada1998a,Radue2013a}. For T$<$T$_c$ VO$_2$ is insulating and has a monoclinic crystal structure, while for T$>$T$_c$  it is metallic and as a rutile lattice structure \cite{Morin1959}. The lower symmetry of the monoclinic phase is characterized by the dimerization of the V$^{+4}$ ions along the c-axis of the rutile phase\cite{Zylbersztejn1975a}. This change in the lattice structure, to which the electron-electron interactions contribute, results in the opening of a band gap in the band structure, illustrated in Fig.\ref{bandandlattice}. \cite{Goodenough1971}. While a complete description of the transition mechanism has not yet been achieved, it is generally accepted that the MIT in VO$_2$ is due to the interplay of a Mott-Hubbard electronic instability in which the electron-electron interactions play the critical role, and a Peierls instability of the lattice.\cite{Cocker2012a,Tao2012a,Kubler2007a,Okazaki2004a}

There are numerous potential applications for VO$_2$.\cite{Imada1998a,Wang2012d,Seo2010,Lappalainen2008} Implementation of such VO$_2$-based new technologies requires the ability to tailor the MIT properties to the demands of the particular application. For example, several studies demonstrated that the critical temperature of a thermally-induced MIT can be adjusted by doping VO$_2$ films, or by applying pressure along the rutile c-axis. \cite{Tselev2010a,Kikuzuki2010c,West2008}. The MIT characteristics also exhibit dependence on the choice of the substrate material. The substrate material can affect the VO$_2$ thin films characteristics [eg. due to strain]. \cite{Abreu2012a,Lu2008a,Kittiwatanakul2011a,Kikuzuki2010b,Aetukuri2013a}. In particular, our previous studies have shown significant differences between the MIT critical temperatures for VO$_2$ samples grown on TiO$_2$ $[011]$, SiO$_2$ (quartz) and c-Al$_2$O$_3$ under identical deposition conditions \cite{Radue2013a}.
In the experiments reported here, we have focused our investigations to the effects of a substrate (material and microstructure) on the ultra-fast photo-induced MIT in two VO$_2$ thin film samples, specifically grown on TiO$_2$ $[011]$ and c-Al$_2$O$_3$. For these measurements the MIT was induced by a strong ultrashort ($\sim$100 fs) laser pulse, and the optical properties of the films were probed by a much weaker probe pulse, arriving with a controlled delay. Following the methodology, proposed by Cocker et al.\cite{Cocker2012a}, we have observed and mapped out a range of values of the pump laser fluence corresponding to distinct phases of the photo-induced MIT at various sample temperatures. We then analyzed the possible connection between the two film's microstructure and observed differences in MIT dynamics.

\section{DIFFERENCES IN THE VO$_2$ FILMS STRUCTURE FOR THE DIFFERENT SUBSTRATES}
In these studies we used two VO$_2$ thin films, produced using the same method (Reactive bias ion target beam deposition \cite{West2008a}) on two different crystalline substrate materials: one 80 nm-thick VO$_2$ film was grown on a 330 $\mu$m c-Al$_2$O$_3$ substrate, and the other 110 nm-thick VO$_2$ film  was grown on a 500 $\mu$m TiO$_2$ (011) substrate. 

Both VO$_2$ films  were characterized by X-ray diffraction (XRD), as outlined in Ref[7] and Ref[10]. The XRD measurements of the sample grown on Al$_2$O$_3$ substrate show the film orientation in the (020) direction with six-fold in-plane symmetry, corresponding to three possible orientations of the VO$_2$ grains within the film.
 The VO$_2$ film grown on TiO$_2$ substrate exhibits a mono-crystalline structure, with a single XRD peak corresponding to a highly-strained monoclinic structure approaching a rutile (011) plane. Such high degree of strain is due to the larger lattice constant of TiO$_2$ substrate resulting in clamping of epitaxial VO$_2$ films\cite{Lu2008a,Muraoka2002a}.

Unsurprisingly, these structural differences have an effect on the characteristics of the thermo-induced MIT\cite{Radue2013a}. Namely, the MIT at the VO$_2$/Al$_2$O$_3$ sample occurred at higher transition temperature (T$_c$=341 K) and had broader width (T$_c$=26 K), compared to the  VO$_2$/TiO$_2$ (T$_c$= 310 K, T$_c$=9 K ). This is possibly due to a wider distribution of grain sizes and strain in the film that has been observed in similar samples in previous papers.\cite{Radue2013a,Aliev2006a,Khakhaev1994a, Brassard2005a}.  

We measured the DC conductivity with applying a 4-point probe technique and found the DC conductivity of VO$_2$ on Al$_2$O$_3$ at 370 K to be $2.24 \times 10^5$ S/m, while the DC conductivity of VO$_2$ on TiO$_2$ at 340 K was found to be $3.03 \times 10^5$ S/m. We also evaluated the penetration depth  at 800 nm for both films. For the film grown on Al$_2$O$_3$ we looked at the reflection and transmission with a cw laser, which gave us a penetration depth of $\delta=294$ nm, where  $I=I_o\times 10^{-x/\delta}$. The TiO$_2$ substrate did not have a polished backside, so we used ellipsometry to measure its optical constants at a range of wavelengths between 420 nm to 749 nm. We took this data and extrapolated to get the real and imaginary parts of the optical constants, and determined a penetration depth at 800 nm optical wavelength to be $\delta=255$ nm. 
\section{EXPERIMENTAL APPARATUS}

\begin{figure}[h!tbp]
\includegraphics[width=\textwidth]{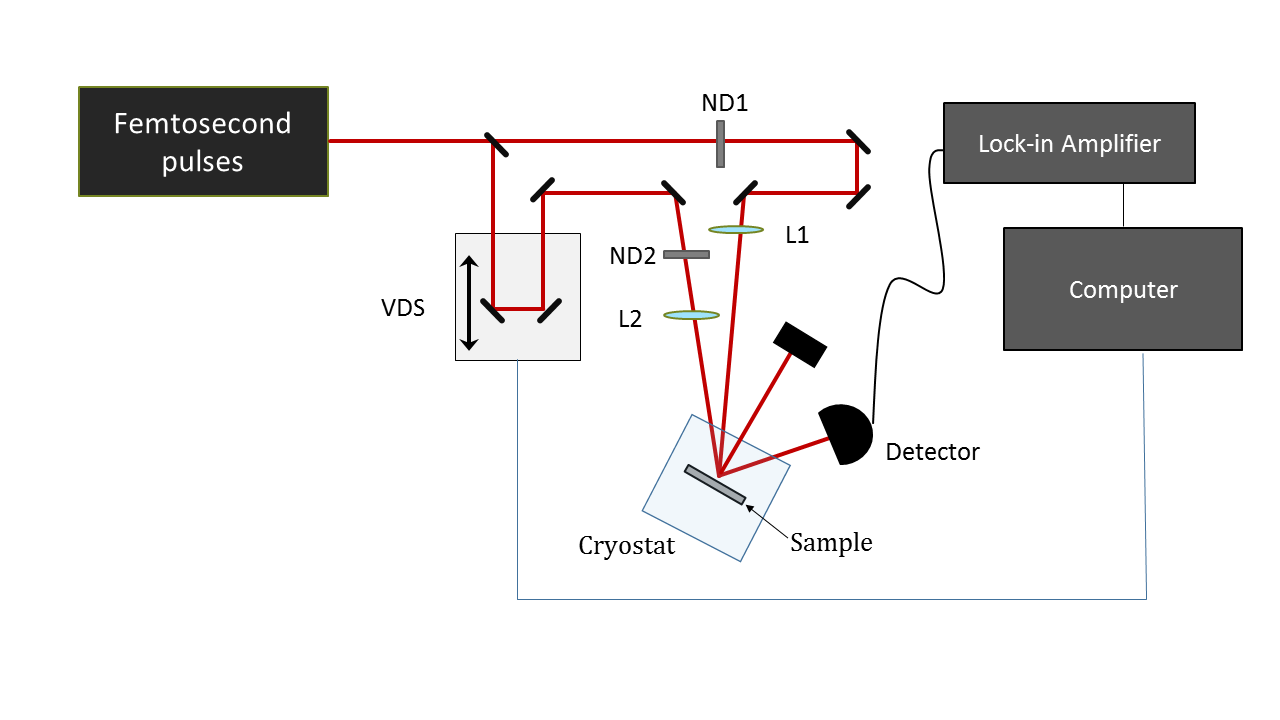}
\caption{Schematics of the optical pump-probe experimental setup. The output of the amplified ultrafast laser is split into weak probe and strong pump beams using a (20/80) beamsplitter (BS). The energy of the pump pulses is controlled by the variable neutral-density filter ND1, while the probe beam was sent through a computer-controlled variable delay stage (VDS) and further attenuated using ND2 (OD=3.0). The probe and pump beams were focused on the same spot at the surface of the sample, placed inside the cryostat, using 250-mm and 500-mm lenses, correspondingly.  The reflected probe power was measured by the photodetector (PD), and further analyzed using a lock-in amplifier.}
\label{setup}
\end{figure}
For our experiments we used the pump-probe configuration shown in Fig.\ref{setup}. Our Ti-sapphire laser system produced 100 fs pulses at a wavelength of 800 nm, which was used for both pump and probe pulses.  The size of the focal spot for the pump beam (180 $\mu$m diameter) was adjusted to be approximately twice larger than that of probe beam (90 $\mu$m diameter) to ensure that the probed area was exposed to relatively uniform pump fluence. The sample temperatures were controlled using an optical cryostat. 

In all measurements, reported below, the relative change in reflection $\Delta R/R$ is defined as $\frac{R_o-R(\tau)}{R_o}$, where R($\tau$) is the power of the measured reflected probe beam as the  function of the delay $\tau$ between the pump and the probe pulses, and $R_o$ is the probe reflected power in the absence of the pump beam. 
\section{MEASUREMENTS OF MIT TEMPORAL EVOLUTION}
\begin{figure}[!htbp]
\includegraphics[width=\textwidth]{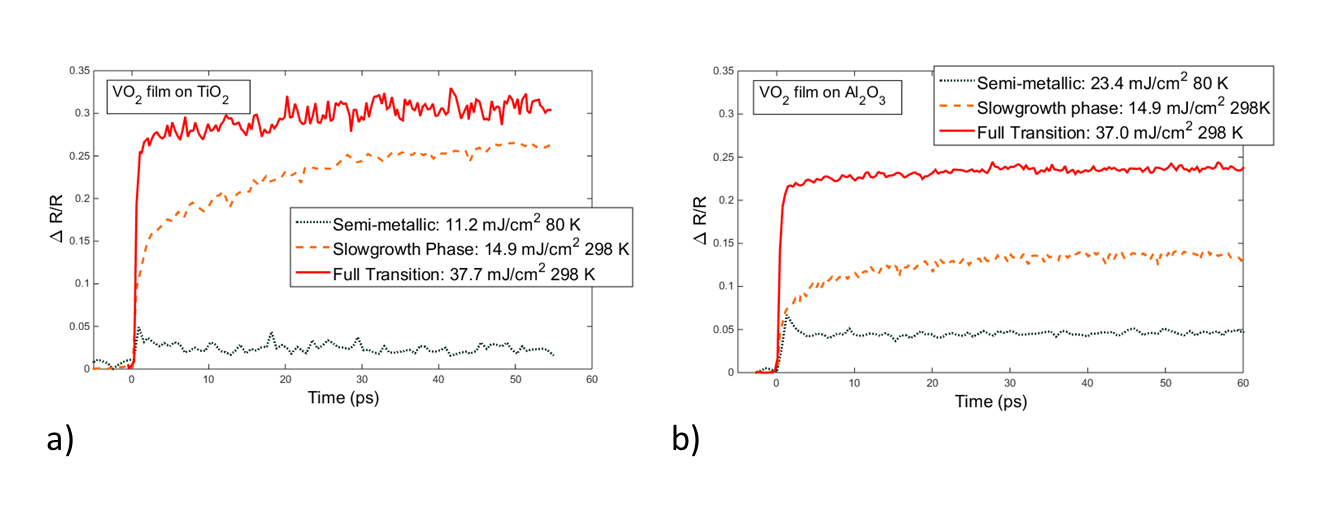}
\caption{Sample time-dependent changes in the probe reflection corresponding to various dynamical phases of MIT, measured for (a) VO$_2$/TiO$_2$ and (b) VO$_2$/Al$_2$O$_2$ samples.  The pump pulse hits the sample at zero time, and positive delay times correspond to the probe beam interacting with the sample after the pump beam.}
\label{traces}
\end{figure}
Overall, the temporal evolution of the photo-induced transition in both samples followed similar trends, as illustrated in Fig. \ref{traces}.  We varied the pump pulse energy and the temperature we held the sample, keeping the initial temperature well below the T$_c$. For both films we observed three distinctive dynamical MIT regimes. Similar behavior has been previously reported by Cocker et al\cite{Cocker2012a} in their studies of the THz probe transmission evolution in a photo-induced MIT in a VO$_2$ film deposited on Al$_2$O$_3$. 
At sufficiently high fluences, we observed a very fast ($<1$ ps) change of the probe reflectivity from the value corresponding to the insulating VO$_2$ state to the fully metallic VO$_2$ state, that implies that the whole VO$_2$ films quickly transitioned from the insulating to the fully metallic state (“full-MIT” phase). We characterize the minimum pump fluence value required to achieve this full transition as $\Phi_{full-MIT}$. 
\begin{figure}[h!tpb]
\includegraphics[width=\textwidth]{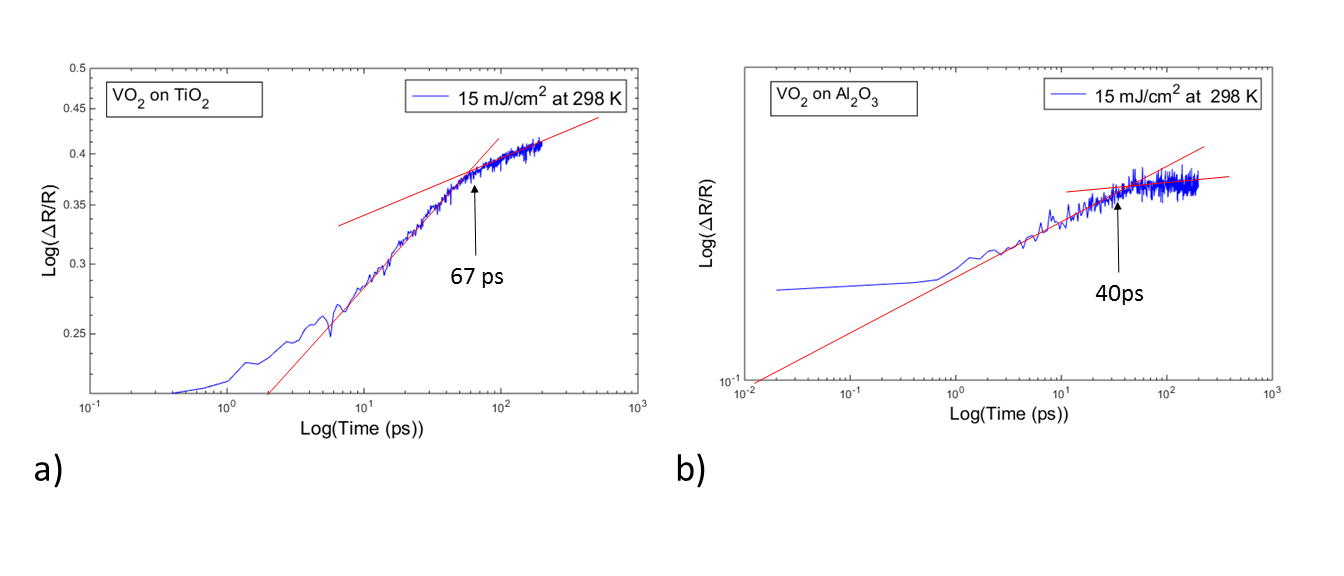}
\caption{“Slow growth” of relative reflectivity with time, measured for (a) VO$_2$/TiO$_2$ and (b) VO$_2$/Al$_2$O$_2$ samples with pump fluence of 15 mJ/cm$^2$ at 298 K, using a log-log scale (“slow growth” phase data in Fig.\ref{traces}. Log-log scale clearly shows two time scales of growth for both films. For the VO$_2$/Al$_2$O$_3$ sample, the initial faster reflectivity growth (time constant 7.4$\pm$ 0.3 ps) for the first 40 ps was followed by much slower growth with time constant of 33$\pm$ 8 ps. For the VO$_2$/TiO$_2$ film the switch occurred at approximately 67 ps, from the time constant 6.3$\pm$0.2 ps to 15.0$\pm$ 0.6 ps.}
\label{loglog}
\end{figure}
 \begin{figure}[!htbp]
\includegraphics[width=\textwidth]{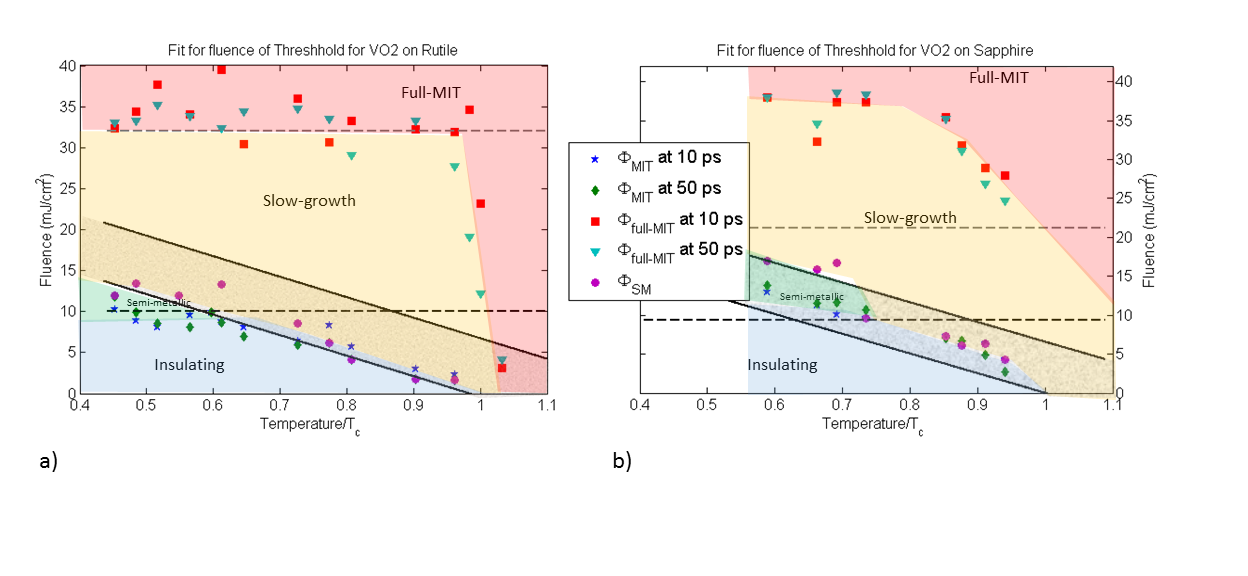}
\caption{Threshold measurements for the onset and for the full MIT of VO$_2$ thin films grown on a)TiO$_2$ and b)Al$_2$O$_3$. Blue stars/Green diamonds correspond to the threshold fluence needed to see a 2$\%$ rise in reflectivity at 10 ps and 50 ps, correspondingly. The red squares and cyan triangles correspond to the fluence required to reach the reflectivity value for a fully metallic film in the first 10 ps and 50 ps, correspondingly. The purple circles indicate the boundary where the ’slow-growth’ phase starts $\Phi_{SM}$. To define this boundary, we took several measurements of the probe reflectivity $\Delta R/R$ as a function of time for the first 20 ps at a range of pump fluences, fitted their linear slopes, and then found fluence values at which the slope becomes indistinguishable from zero within the measurement uncertainty.  The dashed lines are the calculated photo-excited electron densities at the front and the back of the films. The shaded region between the solid lines is calculated fluence needed to bring the films up to the transition temperature.}
\label{Phasediagrams}
\end{figure}
For intermediate values of the pump fluence below $\Phi_{full-MIT}$, the film reflectivity also changed immediately after the pump pulse, but this change did not reach the value corresponding to the fully metallic VO$_2$, implying that the resulting state of the film contains both metallic and insulating regions. Moreover, after this initial sub-picosecond change the relative reflectivity continued to grow at a much slower rate for a few hundreds of ps (“slow growth” phase), indicating further coarsening of the metallic component of the VO$_2$ film.  Interestingly, this dynamical behavior is consistent with the slow coarsening of metallic puddles within the insulating matrix, observed in the thermo-induced MIT.\cite{Qazilbash2006a,Liu2013,Abreu2014} Fig. \ref{loglog} shows the change in both samples reflectivity $\Delta R/R$ at 15 mJ/cm$^2$ on a Log-Log scale to characterize the rate of this evolution more clearly. It is easy to see the two distinct time constants of this slow growth for both films, with two distinct switching times. In general, rate of slow growth decreased at the lower values of the pump fluence. 

At lower values of the pump fluences, the time evolution of the film became dependent on the temperature of the sample. At higher temperatures if the pump fluences is insufficient to reach the slow growth regime, the reflectivity of the sample after the first few ps returned to its insulating value, indicating no long-term changes in the VO$_2$ state. However, at the lower temperatures (below 250 K for VO$_2$/Al$_2$O$_3$ and below 220 K for VO$_2$/TiO$_2$) we clearly observed an intermediate “semi-metallic” phase, as shown in Fig \ref{traces}. For this regime, which occurs at lower fluences than the slowgrowth phase, the pump pulse induced the initial few percent change in relative reflectivity, but then it remained constant, implying no further changes in the VO$_2$ film phase distribution. 
While we observed a qualitatively similar photo-induced MIT dynamics for the two VO$_2$ films grown on TiO$_2$ and on Al$_2$O$_3$ substrates, the structural differences between the two films clearly affected the experimental conditions for realization of each MIT phases.  To highlight these differences, Fig. \ref{Phasediagrams} shows the experimentally measured values of the fluence thresholds between the phases: $\Phi_{full-MIT}$, the fluence required to induce a fast full transition to the metallic stage; $\Phi_{SM}$, the fluence sufficient to induce the partial MIT, followed by the slow growth of the metallic component; and $\Phi_{MIT}$, the minimum pump fluence required to produce any phase transition in the insulating VO$_2$ film. For consistency each threshold value was measured for 10 ps and 50 ps delay between the pump and the probe pulses. Also, to take into account the difference between the critical temperatures for the two VO$_2$ samples, for the phase diagram in Fig. \ref{Phasediagrams} we used the relative temperature T/T$_c$ for each sample.

The most noticeable difference between the two samples was observed in the temperature dependence of the full-MIT threshold $\Phi_{full-MIT}$ . For the VO$_2$ film on Al$_2$O$_3$ the values of $\Phi_{full-MIT}$ increased slowly as the sample temperature dropped. At the same time, for the VO$_2$/TiO$_2$ sample the entire full MIT boundary $\Phi_{full-MIT}$ remained constant for all the temperatures T/T$_c < 0.95$, after the sharp drop around T$_c$. Remarkably, the measure values for in the two samples were rather similar (around 32-34 mJ/cm$^2$ for T/T$_c\sim 0.6)$ despite their structural differences of the films.

The other two threshold values, corresponding to a partial MIT displayed similar temperature dependences for both samples: at lower sample temperatures it required higher pump power to induce the same change in the sample. However, the exact values for $\Phi_{MIT}$ and $\Phi_{SM}$ were higher for the VO$_2$/ Al$_2$O$_3$ sample. We also observed the separation of the slow-growth phase and the semi-metallic phase for temperatures below T/T$_c \leq 0.7$ for the both films. 

It is important to note that the MIT dynamics for VO$_2$/Al$_2$O$_3$ sample observed in the current experiment using all-optical pump-probe detection, was qualitatively similar to that for a similar VO$_2$/Al$_2$O$_3$ sample studied by Cocker et al.\cite{Cocker2012a} using 800 nm optical pump pulses and broadband THz probe pulses. Their reported values for threshold fluences were 2-3 times lower than those shown here in Fig. \ref{Phasediagrams}.  However, other works have shown higher threshold values for the onset of MIT, around 7 mJ/cm$^2$, and show comparable behavior at higher fluences. \cite{Hilton2007a, Cavalleri2001c}

\section{MECHANISMS OF ULTRAFAST MIT}
Accurate theoretical description of the fast dynamics of the electron system in VO$_2$ is a very challenging problem. It requires a proper description of the strong electron-phonon coupling and of the strong electron-electron correlations in the insulating phase, and of the interplay between these correlations and the lattice structure dynamics. However, the analysis of the experimentally measured time evolution of the VO$_2$ reflectivity after the pump pulse may allow us to infer qualitatively the dominant processes that govern the dynamics in the VO$_2$ films that we have studied and, in particular, how these processes differ between VO$_2$ films deposited on TiO$_2$ and on Al$_2$O$_3$.

After the pump pulse the reflectivity, $\Delta R/R$, of the VO$_2$ film, that is initially in the insulating phase, increases on a very short time scale\cite{Kubler2007a}, shorter than our time resolution $\sim$ 0.5 ps, as shown in Fig. \ref{traces}. The change in reflectivity soon after the pump pulse was shown to be directly proportional to the pump fluence,\cite{Kubler2007a} and therefore can be attributed to the almost instantaneous excitation of electrons from the valence band to the conduction band of the monoclinic phase of VO$_2$ by the pump pulse. 

As shown clearly in Fig. \ref{traces}, and described in the previous section, the behavior on time scales longer than 1 ps strongly depends on the pump fluence  and the temperature of the sample\cite{Cocker2012a}.

For sufficiently high values of the fluence $\Phi \geq \Phi_{full-MIT}$, the reflectivity increased to the values matching that of the fully metallic VO$_2$ film\cite{Radue2013a}. This indicates that for this regime the population of electron-hole excitations created by the pump pulse was large enough to cause a structural deformation\cite{Kubler2007a,Biermann2005a} throughout the whole VO$_2$ film and nearly instantaneously drove it from the monoclinic to the rutile phase. The rutile phase is metallic and therefore the excited electrons have no states in the valence band to which they can decay to. For this reason, the film remains metallic until the lattice cools down to temperatures below T$_c$ causing the reversed transition from rutile to monoclinic. This relaxation process is slow and takes place on time scales much longer ($>$ a few ns) than the time interval (few hundred ps) considered in this experiment.

In the limit of weak pump pulses (below a threshold value $\Phi_{MIT}$), the observed change in reflectivity did not last much beyond the duration of the pump pulse, quickly returning to its original value. This can be attributed to the fast relaxation of the particle-hole excitations created by the pump laser: since the number of electron-hole excitations is not high enough to induce a coherent lattice deformation and to drive a noticeable fraction of the VO$_2$ film from the monoclinic insulating phase to the rutile metallic phase, electrons decay quickly from the conduction to the valence band of the monoclinic structure, quickly restoring the original value of the measured reflectivity.

The time evolution of the reflectivity is more complex at intermediate values of the fluence $\Phi_{MIT}<\Phi<\Phi_{full-MIT}$. For these values, the reflectivity kept increasing after the almost instantaneous step. We attribute this behavior to the fact that for $\Phi_{MIT}<\Phi<\Phi_{full-MIT}$ the population of electron-hole excitations created by the pump pulse is large enough to trigger the coherent structural deformation that drives the lattice from the monoclinic phase to the rutile phase in some regions of the sample but not in the whole sample. The rutile regions initially form, then heat up the surrounding monoclinic regions and drive them to the rutile phase, as well. This process leads to coarsening of the metallic regions, and may be responsible for the observed continuing increase of the reflectivity on time scales of the order of tens of picoseconds. It also may explain the existence of two timescales in the slow growth regime, shown in Fig. \ref{loglog}: at shorter times the change in reflectivity can be dominated by lattice vibrations that can persist up to 100 ps after the initial pump pulse excitation\cite{West2008}. At later times, heat diffusion mechanisms dominate the transition dynamics. 

In this model we expect the values of $\Phi_{MIT}$ to decreases as the sample temperature approaches the critical temperature of the thermo-induced transition, since a lower number of electron-hole excitations is necessary to drive the structural phase transition. It is also expected that the semi-metallic phase should appear at the lower temperatures (T/T$_c < 0.7$ for both samples) and lower pump fluences $\Phi_{MIT}<\Phi<\Phi_{SM}$, corresponding to a situation in which the power is sufficient to trigger the MIT transition in isolated regions, but not enough to induce the growth of the metallic regions to the rest of the sample. 

We now discuss how the measured structural differences between the two VO$_2$ films may affect the dynamics of the photo-induced MIT.  The large measured strain induced by the TiO$_2$ substrate indicates that the equilibrium monoclinic VO$_2$ film is already deformed toward rutile structure, thus making the structural transition easier.  This, and the more ordered mono-crystalline structure of the VO$_2$/TiO$_2$ samples  seems to favor the formation of the more uniform nucleation sites during MIT, as evident by a the narrower thermally-induced MIT in this sample, happening at lower critical temperature. The significantly broader width of the MIT transition in the VO$_2$/Al$_2$O$_3$ sample implies the broader distribution\cite{Aliev2006a} of the metallic nucleation cluster sizes compared to VO$_2$/TiO$_2$ samples, as the connection between the width of the thermally-induced MIT and the structure of the film have been demonstrated previously.\cite{Radue2013a,Aliev2006a,Khakhaev1994a, Brassard2005a}

These structural differences may also play a critical role in explaining the observed differences in temperature dependence of the full MIT threshold $\Phi_{full-MIT}$ for the two samples. The absence of the temperature sensitivity of $\Phi_{full-MIT}$ for VO$_2$/TiO$_2$ sample indicates that once the critical density of the photo-electrons is reached, the whole film undergoes the MIT uniformly, independently of its original temperature.  While there is no direct proof, this observation is consistent with a more strained and more ordered monocrystalline structure of the sample, as well as the sharper thermally-induced MIT transition.  In contrast, the less ordered structure of the VO$_2$/Al$_2$O$_3$ sample may give rise to stronger local variations of the critical temperature throughout the sample (as indicated by a much wider thermally-induced MIT transition). This non-uniformity can then be reflected in stronger local variations in the critical density of particle-hole excitations necessary to induce the coherent lattice distortion and to drive the structural transition, resulting in the critical density of electron-hole excitations necessary to drive the full MIT in the whole sample to be temperature-dependent. 

We can attempt to estimate the minimum pump fluence values required to excite the critical density of photoelectrons to match the free electron density in the metallic phase of the VO$_2$ film in equilibrium (above the critical temperature). First, we estimate the electron density in the metallic VO$_2$ films from the measured dc electrical conductivity, using the Drude-Smith model\cite{Cocker2010b}:
	\begin{equation}
	\sigma_{DC}=\frac{ne^2\tau_{DS}}{m^*}\times (1+c)
	\label{DCconductivity}
	\end{equation}
where $n$ is the electron density, $\tau_{DS}$ is the scattering time, $m*$ is the effective mass of the electrons, and $c$ is the correlation parameter, such that $c=0$ corresponds to free Drude electrons, and $c=-1$ corresponds to fully localized electrons. Using the previously established values for $\tau_{DS}$, $m∗$, and $c$,\cite{Cocker2010b} the estimated value of electron densities are $3.2 \times 10^{21}$cm$^{-3}$ for the VO$_2$/TiO$_2$ sample, and $2.35 \times 10^{21}$cm$^{-3}$ for the VO$_2$/Al$_2$O$_3$ sample. 

Second, we approximate the number of photo-electrons by the number of photons absorbed; although in this way we clearly overestimate the photoelectron density, this provides an order of magnitude estimates of the required fluences. Following a similar procedure, described in Ref.[9], we compute the values of the MIT thresholds $\Phi_{MIT}$  and $\Phi_{full-MIT}$ based on reaching the critical density of the photoelectrons at the front of the film (using full pump laser fluence) and in the back of the film (using the attenuated pump laser fluence due to reflection and absorption inside the film). Using the measured values of the penetration depth of the 800-nm optical beam, we estimated the threshold fluence needed to photo-excite the metallic electron density at the front of the VO$_2$/TiO$_2$ film to be approximately 10 mJ/cm$^2$, and the fluence required to excite the metallic electron density at the back of the film to be 36 mJ/cm$^2$. These values are shown as dashed horizontal lines in Fig. \ref{Phasediagrams}(a). The higher threshold matches the boundary measured for the full metal-insulator transition.
Analogous calculations for the VO$_2$ film on Al$_2$O$_3$ lead to threshold fluence values of 9.3 mJ/cm$^2$ and 21 mJ/cm$^2$, respectively.  We can see that for Al$_2$O$_3$ both thresholds obtained in this way do not match the $\Phi_{MIT}$ and $\Phi_{full-MIT}$ thresholds obtained experimentally.

Finally, we roughly estimated the minimum pump fluence values required to heat the VO$_2$ films to the thermally-induced phase transition, assuming that all the absorbed optical energy was eventually transferred to heating the sample. The two solid lines in Fig. \ref{Phasediagrams} are due to the finite width of the MIT, as can be seen in Table \ref{table}. The lower solid lines correspond to the minimum pump fluence required to increase the sample temperature to the beginning stage of the thermally-induced MIT (specifically to T=305 K for the VO$_2$/TiO$_2$ sample, and T$_c$ = 325 K for the VO$_2$/Al$_2$O$_3$ sample). The upper lines assumed that the samples reached the fully metallic phase at T=315 K and T=355 K respectively, and included the latent heat required for the phase transition. In these estimates we used the value of heat capacity of the VO$_2$ film to be 3.0 J/cm$^3$ K, and the latent heat of 235 J/cm$^3$.\cite{Berglund1969a} For both films we find that the experimental $\Phi_{MIT}$ threshold matches semi-quantitatively the threshold obtained by estimating the value of the fluence needed to raise the temperature of the VO$_2$ film to T$_c$.
\begin{table}[h!]
\begin{tabular}{|p{6cm}|p{4cm}|p{4cm}|}
\cline{1-3}
     & Film grown on TiO$_2$  & Film grown on  Al$_2$O$_3$   \\ \cline{1-3}
  Average VO$_2$ grain size (from XRD) & Out of plane: $>$13 nm  & Out of plane: 45 nm   \\ \cline{1-3}
  VO$_2$ film thickness& 110 nm & 80 nm   \\ \cline{1-3}
  Thermal Transition Temperature T$_c$ & 310 K  & 340 K   \\ \cline{1-3}
  MIT width $\Delta$T$_c$ & 9 K  & 26 K  \\ \cline{1-3}
  DC Conductivity  & at 297 K $7.01\times 10^3$ 
  
  at 340 K $3.03\times 10^5$
    & at 296K $1.71\times 10^2$

    at 369 K $2.24\times 10^5$\\ \cline{1-3}
  Penetration depth at 800 nm  & 255 nm  & 294 nm  \\ \cline{1-3}
  Slowgrowth threshold fluence $\Phi_{MIT}$  & 1.2 mJ/cm$^2$ at 300 K
  
  4.2 mJ/cm$^2$ at 250 K 
  
  13.3  mJ/cm$^2$ at 190 K
    & 6.1 mJ/cm$^2$ at 300 K
    
    10.6  mJ/cm$^2$ at 250 K
    
    16.7  mJ/cm$^2$ at 200 K
     \\ \cline{1-3}
  Pump fluence required induce the full transition   & 27.8 mJ/cm$^2$ at 300 K

  33.4 mJ/cm$^2$  at 250 K

  32.4 mJ/cm$^2$ at 190 K
    & 31.9 mJ/cm$^2$ at 300 K

    37.4 mJ/cm$^2$ at 250 K

    37.9  mJ/cm$^2$ at 190 K
     \\ \cline{1-3}
  Semi-metallic stage emergence conditions & 8.5 mJ/cm$^2$ at 220 K  & 12 mJ/cm$^2$ at 250 K \\ \cline{1-3}
\end{tabular}
\label{table}
\caption{Summary of basic properties of the two VO$_2$ samples.}
\end{table}
\section{CONCLUSION}
In conclusion, we have investigated the role of substrate-induced microstructure on the ultrafast dynamics of the photoinduced metal-insulator transition in VO$_2$ films, using optical pump-probe techniques. In particular, we have identified the characteristic patterns of the MIT dynamics that depend on pump laser energy and the temperature of the film, and measured the threshold pump fluence values that characterize the onset and full MIT between such behaviors. We found that two VO$_2$ samples - one grown on sapphire substrate, and the other grown on rutile substrate, require rather similar threshold laser fluence values, despite a rather significant (30 K) difference in the critical temperature for the thermally-induced MIT. We also identified several important differences, summarized in Table 1. In particular, for the VO$_2$ film on TiO$_2$ we found almost no temperature dependence in the values of fluence threshold for the full MIT, while for the VO$_2$/Al$_2$O$_3$ sample this threshold clearly decreased as T approaches T$_c$. This difference may be linked to the large lattice strain and differences in width of the thermo-induced transition, which we suspect is linked to the variance of the grain size of the film. Understanding the changes in the transition dynamics of films on difference substrates will potentially allow for the selection of desirable traits in the MIT, which is important for future VO$_2$-based technologies.
	
\begin{acknowledgments}
This work was supported by NSF, DMR-1006013: Plasmon Resonances and Metal Insulator Transitions in Highly Correlated Thin Film Systems, and the NASA Virginia Space Grant Consortium. We also acknowledge support from the NRI/SRC sponsored ViNC center and the Commonwealth of Virginia through the Virginia Micro-Electronics Consortium (VMEC), and Jeffress Trust Awards program in Interdiciplinary Research.
\end{acknowledgments}
\bibliography{C:/Users/Elizabeth/Documents/library}

\end{document}